\documentclass[11pt]{amsart}
\usepackage{geometry}                                   \usepackage{graphicx}
\usepackage{amssymb}
\usepackage{hyperref}
\usepackage{booktabs}
\usepackage{subfig}
\usepackage{engord}
\usepackage{amsaddr}
\usepackage{url}
\usepackage[backend=biber,sorting=none,style=numeric-comp,urldate=long,maxcitenames=10,maxbibnames=9]{biblatex}
\usepackage{todonotes}
\usepackage[normalem]{ulem}

\usepackage[nomargin,inline,marginclue,draft]{fixme}
 \fxsetup{theme=color, mode=multiuser} 
 \FXRegisterAuthor{pb}{apb}{\color{blue}PB} 
 \FXRegisterAuthor{mp}{avs}{\color{red}MP}

\usepackage{todonotes}
\usepackage{xargs} 
\newcommandx{\pbtodo}[2][1=]{\todo[linecolor=red,backgroundcolor=red!25,bordercolor=red,#1]{\tiny PB:  #2}}
\newcommandx{\mptodo}[2][1=]{\todo[linecolor=blue,backgroundcolor=blue!25,bordercolor=blue,#1]{\tiny MP: #2}}

\usepackage[acronym]{glossaries}

\newacronym{esco}{ESCO}{European standard classification of occupations}
\newacronym{sors}{SORS}{Statistical office of the Republic of Slovenia}
\newacronym{isco}{ISCO}{International Standard Classification of Occupations}

\title{Occupation similarity through bipartite graphs}
\author{Pavle Boškoski, Matija Perne}
\address{Jožef Stefan Institute, Jamova cesta 39, 1000 Ljubljana, Slovenia}
\email{pavle.boskoski@ijs.si}
\author{Tjaša Redek}
\address{Faculty of Economics, University of Ljubljana,\\ Kardeljeva ploščad 17, 1000 Ljubljana, Slovenia}
\author{Biljana Mileva Boshkoska}
\address{Jožef Stefan Institute, Jamova cesta 39, 1000 Ljubljana, Slovenia}
\address{Faculty of Information Studies in Novo mesto, Ljubljanska cesta 31b, 8000 Novo mesto, Slovenia}

\begin{document}
\begin{abstract}
Similarity between occupations is a crucial piece of information when making career decisions. However, the notion of a single and unified occupation similarity measure is more of a limitation than an asset. The goal of the study is to assess multiple explainable occupation similarity measures that can provide different insights into inter-occupation relations.
Several such measures are derived using the framework of bipartite graphs.
Their viability is assessed on more than 450,000 job transitions occurring in Slovenia in the period between 2012 and 2021.
The results support the hypothesis that several similarity measures are plausible and that they present different feasible career paths.
The complete implementation and part of the datasets are available at \url{https://repo.ijs.si/pboskoski/bipartite_job_similarity_code}.
\end{abstract}
\maketitle

\section{Introduction}

The labour landscape is constantly changing and the labour market is getting more dynamic~\cite{10.2767/653528}, while the ongoing pandemic is hastening these changes~\cite{10.2767/123325}.
Navigating in the maze of ever-changing career options is challenging in these circumstances.
Accurate and timely information is becoming essential for selecting the optimal career path.
A crucial piece of information that can help improve one's decision making process is occupation similarity.
It can serve at least two purposes.
It can directly map a possible career evolution plan~\cite{10.3368/jhr.53.2.0814-6556R2,arXiv:2109.02554,10.5220/0005702302700277}, and it can help identify risky occupations that are in decline and from which it is difficult to find a viable career path~\cite{10.1371/journal.pone.0254722,10.1080/04353684.2021.1884497}.
The latter is very important for policy makers in the preparation of active labour market policy plans, particularly in the realms of retraining and life-long learning~\cite{10.1126/sciadv.aao6030,klaus_schwab_zahidi_2020}.
In such a complex setting, it should not be expected that a single similarity measure will provide a generally applicable solution.
There may be several explainable occupation similarity measures, and the job seeker's preferences or other factors like career status may determine which one of them is the most applicable.

Occupation similarity can be directly estimated from the observed job transfers occurring within a certain economy~\cite{10.1371/journal.pone.0254722}.
It, however, requires a large enough amount of accurate micro(personal)-level job transfer data, which is rarely available.
In the cases where such data is available, the sample size is typically in the order of several thousand records, which is very limiting.
As a result, the generality of results drawn in such a manner is questionable.

Another way of estimating the similarity between occupations is to exploit the available structured data sets that provide links among occupations, skills, tacit knowledge and education.
There are several such structures, the most prominent ones being the US Occupational Information Network (O*Net)~\cite{onet} and the \gls{esco}~\cite{esco}.

The presented work takes advantage of both possibilities for determining the occupation similarity.
The ontology structure \gls{esco} that provides a link between occupations, skill groups and skills is used as the structured data set.
The job transfers are observed in the Statistical Register of Employment - SRDAP available at the \gls{sors}.
It is a set of micro level data describing each individual movement in the labour market in Slovenia.
This data set consists of more than 450,000 job transfers spanning the period between 2012 and 2021.
The goal is to propose a computationally efficient occupation similarity measure that uses \gls{esco} data as a source and to validate it using the vast amount of available micro data. 

To the authors' best knowledge, all of the existing studies of similarity between occupations are based on a pre-defined structured data set such as O*Net or \gls{esco}.
This is mainly due to the limited access to micro-level data.
Since the link between two occupations in a structured data set are their corresponding sets of skills, the main challenge is to quantify this relation.
The approaches used belong in two general groups.
Some are based on algorithms quantifying the similarity and diversity of sets.
Among these, the most prominent examples include Jaccard similarity~\cite{arXiv:2109.02554,10.3390/proceedings2021074015}, proximity through revealed comparative advantage~\cite{10.1109/BigData47090.2019.9005967,10.1126/science.1144581}, and asymmetric measures~\cite{1479155}.
The other studies employ the techniques from natural language processing for quantifying skills similarity, at least to some extent.
Typical examples are Nesta~\cite{nesta} based on BERT neural network analysing the skills descriptions and Australian job recommendation employing classifiers~\cite{10.1371/journal.pone.0254722}, among others.

It is important for the models to be explainable~\cite{10.1145/3236386.3241340} and computationally efficient.
If the inner workings of the proposed similarity measures are straightforward, they can be easily presented to the end users, either job seekers or labour policy makers.
Consequently, the users can easily choose the measures that are applicable to their circumstances, such as their wishes, career status, etc.

Since the main sources of information are the sets of occupations and the sets of skills, approaches from the framework of bipartite graphs are also applicable to this problem~\cite{10.1016/j.dss.2012.09.019}.
These techniques are readily applied in the field of recommendation systems~\cite{10.1103/PhysRevE.76.046115,10.1109/ICACIA.2008.4770004,10.1093/gigascience/giy014,10.1073/pnas.1000488107} and complex network analysis~\cite{10.1016/j.physa.2010.11.027,10.1103/PhysRevE.80.046122}.
Particularly promising are the techniques similar to collaborative filtering~\cite{10.1016/j.ipm.2011.03.007}.

Drawing on this vast set of tools, the goal is to show that it is possible to define different explainable occupation similarity measures.
Each one of these measures is relevant to somewhat different career mapping scenarios.
Finally, the measures are validated against the dataset of more than 450,000 real-world observed job transfers.
 
\section{Problem statement}
The \gls{esco} ontology provides two sets of entities that are relevant for determining the similarity of occupations:
\begin{enumerate}
\item Set of occupations 
$
\mathcal{O} = \{ \text{O}_1,\ldots,\text{O}_m\}
$
\item Set of skills
$
\mathcal{S} = \{ \text{S}_1,\ldots,\text{S}_n\}.
$
\end{enumerate}
For each occupation, it specifies which skills are {\em essential} for it and which are {\em optional}.
In addition, \gls{esco} provides a partitioning of $\mathcal{S}$ into blocks $P_j$ of related skills, so that
\begin{equation}
\mathcal{S}_w = \cup_{j=1}^{j=c+k+a+l} P_j.
\end{equation}
Here, $c$, $k$, $a$, and $l$ are the numbers of blocks in $\mathcal{S}_w$ comprising core skills, knowledges, attitudes, and languages, respectively.
We treat these 4 types of skills equally, while \gls{esco} actually provides a hierarchy of differently fine partitions of $\mathcal{S}$ and the available number of levels differs between the types of skills.
We only use $\mathcal{S}$ itself and the next finest partition, which is the partition we label as $\mathcal{S}_w$

\subsection{Skill similarity through bipartite graphs}
The three sets $\mathcal{O}$, $\mathcal{S}$ and $\mathcal{S}_w$ can be used to analyse the relationships among occupations and skills in the context of bipartite graphs, that is, graphs with two disjoint sets of nodes.
Several types of bipartite graphs can be constructed. 

\begin{figure*}[ht]
	\centering
	\subfloat[]{\includegraphics{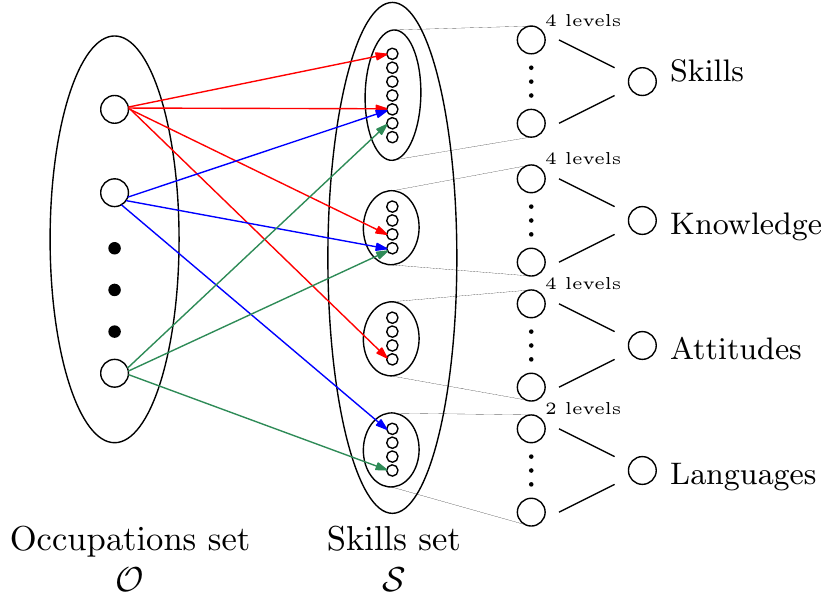}\label{fig:bipartite}}
	\subfloat[]{\includegraphics{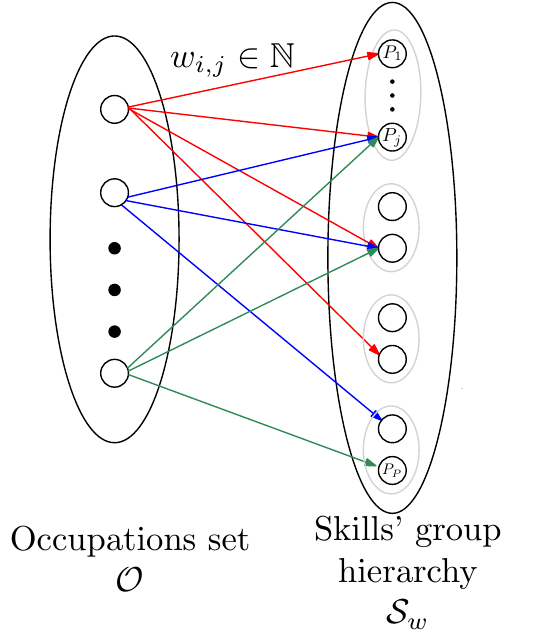}\label{fig:bipartite_weighted}}
	\caption{Bipartite graph representation of the occupation -- skill relations.}
	
\end{figure*}

The simplest approach is to specify an unweighted bipartite graph $G=(\mathcal{O}, \mathcal{S}, E)$, where $E$ is the biadjacency matrix, i.e. a (0,1) matrix of size $|\mathcal{O}| \times |\mathcal{S}|$, the elements of which have the value of 1 if the corresponding vertices are adjacent and 0 otherwise~\cite{Asratian_Denley_Haggkvist_1998}.
A skill and an occupation are taken to be adjacent if and only if the skill is relevant for the occupation.
The resulting bipartite graph is shown in \figurename~\ref{fig:bipartite}.

Another possibility is to use the partitioned set of skills $\mathcal{S}_w$.
The result is a weighted bipartite graph $G_w=(\mathcal{O}, \mathcal{S}_w, E_w)$, where the 
weights $E_w$ are not binary as in $G$.
The weight $E_w^{(i,j)}$ between the occupation $o_i$ and the block $P_j$ is determined from the unweighted bipartite graph $G$.
It equals the number of edges connecting $o_i$ to a skill in $P_j$.
The resulting bipartite graph is shown in \figurename~\ref{fig:bipartite_weighted}.

To utilise the distinction between essential and optional skills, an unweighted bipartite multigraph $G_m=(\mathcal{O}, \mathcal{S}, E_{\mathrm{all}}, E_{\mathrm{ess}})$ can be used.
Here, the biadjacency matrix $E_{\mathrm{all}}$ relates a skill to an occupation if it is either essential or optional, while $E_{\mathrm{ess}}$ only considers essential skills.
By using the partitioned skill set $\mathcal{S}_w$ and distinguishing between essential and optional skills, a weighted bipartite multigraph $G_{wm}=(\mathcal{O}, \mathcal{S}_w, E_{w{\mathrm{all}}}, E_{w{\mathrm{ess}}})$ is obtained.
The weight matrices $E_{w{\mathrm{all}}}$ and $E_{w{\mathrm{ess}}}$ are determined from the matrices $E_{\mathrm{all}}$ and $E_{\mathrm{ess}}$ of the unweighted bipartite multigraph $G_m$ in the same way that $E_w$ is obtained from $E$.
The weight $E_{w{\mathrm{all}}}^{(i,j)}$ equals the number of edges connecting the occupation $o_i$ to a skill in $P_j$, where all the skills in $E_{\mathrm{all}}$ count. The weight $E_{w{\mathrm{ess}}}^{(i,j)}$ is the number of edges connecting the occupation $o_i$ to a skill in $P_j$, where edges represent only essential skills.

\section{Projections of bipartite graphs}
It is possible to infer the relations among the nodes in one set of the bipartite graph from the bipartite relations.
This operation is referred to as projection.
The graph can be projected onto either set of nodes.
To infer the relationships between occupations, the set of skills~$\mathcal{S}$ is projected onto the set of occupations~$\mathcal{O}$.
The result is a 
weighted directed graph $G_d=(\mathcal{O}, D)$, 
where $D$ is the weight matrix with dimensions $|\mathcal{O}| \times |\mathcal{O}|.$ 
The elements of $D$ depend both on the bipartite graph and on the selected projection.

When adapting the projections to accommodate the multigraphs distinguishing between essential and optional skills, we suppose that a person has acquired most of the optional skills of their former job but needs only the essential skills for their possible future job. 
We take into account all essential as well as all optional skills for the source occupation $o_i$, but only pay attention to the essential skills for the target occupation $o_j$.

\subsection{Unweighted bipartite graphs projections}
These projections operate on unweighted bipartite graphs $G=(\mathcal{O},\mathcal{S},E)$ where all the elements of the biadjacency matrix $E$ equal either 0 or 1.
As a result, these projections can be implemented in a computationally efficient manner.
They can be extended to unweighted bipartite multigraphs $G_m=(\mathcal{O}, \mathcal{S}, E_{\mathrm{all}}, E_{\mathrm{ess}})$ as long as all the elements of $E_{\mathrm{all}}$ and $E_{\mathrm{ess}}$ are equal to 0 or 1, enabling the use of the distinction between essential and optional skills.
Among many various projections, two of the most commonly used ones are the overlap weighted projection and collaboration projection.

\subsubsection{Overlap weighted projection}
It is possible to use Jaccard index as a bipartite graph projection.
For the unweighted bipartite graph $G=(\mathcal{O},\mathcal{S},E)$, the projection weight between occupations $o_i$ and $o_j$ is defined as~\cite{10.4135/9781446294413.n28}
\begin{equation}
d_{i,j}^{\mathrm{jacc}'} = \frac{| N(o_i)\cap N(o_j) |}{|N(o_i) \cup N(o_j)|},
\label{eq:jaccard_index_un}
\end{equation}
where $N(o_i)$ is the set of neighbours of node $o_i$ and $|\cdot|$ denotes the set's cardinality.
The set $N(o_i)$ contains all the skills relevant for the occupation $o_i$.
Relation~\eqref{eq:jaccard_index_un} can be considered as the relative overlap of two sets, hence the name.

For the unweighted bipartite multigraph $G_m=(\mathcal{O}, \mathcal{S}, E_{\mathrm{all}}, E_{\mathrm{ess}})$, the definition~\eqref{eq:jaccard_index_un} becomes
\begin{equation}
d_{i,j}^{\mathrm{jacc}''} = \frac{| N_{\mathrm{all}}(o_i)\cap N_{\mathrm{ess}}(o_j) |}{|N_{\mathrm{all}}(o_i) \cup N_{\mathrm{ess}}(o_j)|},
\label{eq:jaccard_index}
\end{equation}
where $N_{\mathrm{all}}(o_i)$ is the set of neighbours of the node $o_i$ as specified in $E_{\mathrm{all}}$ and $N_{\mathrm{ess}}(o_j)$ the set of neighbours of the node $o_j$ defined in $E_{\mathrm{ess}}$.
That is, both the essential and the optional skills of the initial occupation $o_i$ are taken into account, but only the essential skills of $o_j$.

The Jaccard similarity \eqref{eq:jaccard_index_un} is symmetric.
However, in the case of occupation similarity, the ease of transfer between occupations $o_i$ and $o_j$ might not be symmetric. 
In order to accommodate the need for asymmetric similarity, the original definition of the Jaccard similarity~\eqref{eq:jaccard_index_un} can be altered by looking only at the cardinality of target occupation skill,
\begin{equation}
d_{i,j}^{\mathrm{jacc}'''} = \frac{| N(o_i)\cap N(o_j) |}{| N(o_j)|}.
\label{eq:proj_jacc_0}
\end{equation}
The final formula
\begin{equation}
d_{i,j}^{\mathrm{jacc}} = \frac{| N_{\mathrm{all}}(o_i)\cap N_{\mathrm{ess}}(o_j)|}{| N_{\mathrm{ess}}(o_j)|}
\label{eq:proj_jacc}
\end{equation}
is obtained by distinguishing between essential and optional skills in \eqref{eq:proj_jacc_0}.

\subsubsection{Collaboration projection}
The projection was introduced by \citeauthor{10.1103/PhysRevE.64.016132}~\cite{10.1103/PhysRevE.64.016132} for analysis of interactions of authors of scientific articles.
The projection weights are calculated using the formula
\begin{equation}
d_{i,j}^{\mathrm{coll}'} = \sum_{k \in \mathcal{S}} \frac{E^{(i,k)} E^{(j,k)}}{\left(\sum_{l \in \mathcal{O}} E^{(l,k)}\right)-1},
\label{eq:proj_newman_0}
\end{equation}
which turns into
\begin{equation}
d_{i,j}^{\mathrm{coll}} = \sum_{k \in \mathcal{S}} \frac{E_{\mathrm{all}}^{(i,k)} E_{\mathrm{ess}}^{(j,k)}}{\left(\sum_{l \in \mathcal{O}} E_{\mathrm{ess}}^{(l,k)}\right)-1}
\label{eq:proj_newman}
\end{equation}
when the distinction between essential and optional skills is accounted for.

Unlike~\eqref{eq:proj_jacc}, this projection does not assign equal importance to every shared neighbour but assigns more importance to neighbours with fewer neighbours.
In the context of job similarity, sharing a skill that is not shared by many other occupations contributes more than sharing a generic skill.
In this way, this measure looks for niche skills and occupations.

\subsection{Weighted bipartite graph projections}
The graphs $G_w=(\mathcal{O},\mathcal{S}_w, E_w)$ and $G_{wm}=(\mathcal{O}, \mathcal{S}_w, E_{w{\mathrm{all}}}, E_{w{\mathrm{ess}}})$ are weighted. 
Although the above projections are not directly applicable to these, they can be easily extended.

\subsubsection{Generalised Jaccard similarity}
The generalised Jaccard similarity is defined as~\cite{10.1109/TKDE.2018.2876250, 10.1109/ICDM.2010.80}
\begin{equation}
d_{i,j}^{\text{gjacc}'} = \frac{\sum_{k\in\mathcal{S}_w} \min (E_w^{(i,k)},E_w^{(j,k)}) }{\sum_{k\in\mathcal{S}_w} \max (E_w^{(i,k)},E_w^{(j,k)})}.
\label{eq:w_jacc_0}
\end{equation}
Distinguishing between essential and optional skills, it becomes
\begin{equation}
d_{i,j}^{\text{gjacc}} = \frac{\sum_{k\in\mathcal{S}_w} \min (E_{w{\mathrm{all}}}^{(i,k)},E_{w{\mathrm{ess}}}^{(j,k)}) }{\sum_{k\in\mathcal{S}_w} \max (E_{w{\mathrm{all}}}^{(i,k)},E_{w{\mathrm{ess}}}^{(j,k)})}.
\label{eq:w_jacc}
\end{equation}
If the weights $E_w$ are binary, the relations $\min$ and $\max$ correspond to $\cap$ and $\cup$ respectively, i.e.~\eqref{eq:w_jacc_0} reduces to~\eqref{eq:jaccard_index_un}, hence the name.

\subsubsection{Collaborative filtering}
Extending the Newman's collaborative projection~\eqref{eq:proj_newman} for weighted bipartite graphs leads to the concept of collaborative filtering~\cite{10.1109/ICACIA.2008.4770004}.
For each occupation node $o_i$, let $R_{o_i}$ be the weighted node order, i.e. the sum of weights from $o_i$ to all of its neighbours $N(o_i)$,
\begin{equation}
R_{o_i} = \sum_{\forall P_k \in N(o_i)} E_w^{(i,k)}.
\end{equation}
The same can be done for each skill partition $P_k$, with the total adjacency weights (weighted node order) being
\begin{equation}
R_{P_k} = \sum_{\forall o_i \in N(P_k)} E_{w}^{(i,k)}.
\end{equation}
The similarity between occupations $o_i$ and $o_j$ expresses as
\begin{equation}
d_{i,j}^{\text{colf}'} = \frac{1}{R_{o_i}}\sum_{P_k \in \mathcal{S}_w}\frac{E_w^{(i,k)}E_w^{(j,k)}}{R_{P_k}},
\label{eq:recom0}
\end{equation}
or, in the case of a weighted bipartite multigraph,
\begin{equation}
d_{i,j}^\text{colf} = \frac{1}{R_{o_i}}\sum_{P_k \in \mathcal{S}_w}\frac{E_{w{\mathrm{all}}}^{(i,k)}E_{w{\mathrm{ess}}}^{(j,k)}}{R_{P_k}}.
\label{eq:recom}
\end{equation}
The quantities $R_{o_i}$ and $R_{P_k}$ in equation~\eqref{eq:recom} are defined as
$$R_{o_i} = \sum_{\forall P_k \in N(o_i)} E_{w{\mathrm{ess}}}^{(i,k)}$$
and
$$R_{P_k} = \sum_{\forall o_i \in N(P_k)} E_{w{\mathrm{ess}}}^{(i,k)}.$$

Compared to the collaboration projection~\eqref{eq:proj_newman_0}, there are slight differences.
The similarity~\eqref{eq:recom0} is not symmetrical.
Furthermore, the sums for each row of $d_{i,j}$ is one, thus it has a form of normalisation.

\section{Ranking evaluation using different projections}
\label{sec:ranking}
Each of the projections listed in the previous section provides a different similarity value.
Therefore, different ranked lists of similar occupations are available for each starting occupation $o_i$.
The main question is, which one better reflects reality?

The proposed similarity measures are evaluated in three conceptually different ways.
Firstly, the properties of the used bipartite graph projections are studied in more depth.
Secondly, the obtained ranked lists are compared with the ones obtained using an existing occupation similarity measure.
For this purpose, we use the similarity results by Nesta~\cite{nesta}.
Finally, the calculated similarity measures are validated against the actual occupation transfers observed in a society, in this case using the micro-data records from \gls{sors}.

\subsection{Properties of the different projections}
In the context of occupation similarity, projections \eqref{eq:proj_jacc} and~\eqref{eq:proj_newman} each lead to a conceptually-different ordering.
This can be demonstrated by checking the calculated similarity measures for ``plastering supervisor''\footnote{ESCO URI:\url{http://data.europa.eu/esco/occupation/aecb033a-2be8-459c-bbdc-87bbae922894}}, the results of which are shown in \tablename~\ref{tab:proj_comp}.
The overlap weighted projection favours occupations that have the largest number of intersecting skills, hence its name.
Therefore, it provides a ranking that initially explores the local structure and then progresses to the similar branches within the same occupation groups.
As a result, the closest match to ``plastering supervisor'' is the occupation ``plasterer''. 
Collaboration projection favours occupations that share exclusive skills. 
So, the closest matches are the supervisory roles in construction, since they share the skills found in a small number of occupations.

\begin{table}[h]
\centering
\caption[Test]{Similarity distance using unweighted bipartite graph projections for ``plastering supervisor''.}
\label{tab:proj_comp}
\begin{tabular}{ll}
\toprule
            Overlap projection order $d_{i,j}^{\mathrm{jacc}}$&                         Collaboration projection order $d_{i,j}^{\mathrm{coll}}$\\
\midrule
                    plasterer &                     insulation supervisor \\
 building construction worker &                         tiling supervisor \\
        insulation supervisor &                terrazzo setter supervisor \\
       paperhanger supervisor &                    paperhanger supervisor \\
            ceiling installer &                        roofing supervisor \\
       bricklaying supervisor &          construction painting supervisor \\
   terrazzo setter supervisor &  water conservation technician supervisor \\
\bottomrule
\end{tabular}
\end{table}

A similar observation can be performed when using weighted bipartite graph projections.
As shown in \tablename~\ref{tab:weight_proj_comp}, the generalised Jaccard distance projection~\eqref{eq:w_jacc} favours supervisor roles in the construction sector.
The collaborative filtering projection $d_{i,j}^\text{colf}$ \eqref{eq:recom} provides a completely different path. 
It provides broader possibilities that are not predominantly focused on exclusive skills.

\begin{table}[h]
\centering
\caption{Similarity distance using weighted bipartite graph projections for ``plastering supervisor''.}
\label{tab:weight_proj_comp}
\begin{tabular}{ll}
\toprule
Collaborative filtering order $d_{i,j}^{\mathrm{colf}}$&               Generalised Jaccard order $d_{i,j}^{\mathrm{gjacc}}$\\
\midrule
            plasterer &                roofing supervisor \\
      terrazzo setter &  construction painting supervisor \\
         tyre builder &            bricklaying supervisor \\
           bricklayer &                 tiling supervisor \\
    ceiling installer &              carpenter supervisor \\
       door installer &               plumbing supervisor \\
plate glass installer &             insulation supervisor \\
\bottomrule
\end{tabular}
\end{table}

The variation in the results does not mean that one projection is better than the other.
It means that one may shift the focus between a broader and a more targeted skill set through the choice of the similarity measure.
This can be considered as an asset, since the similarity measure can be altered for different scenarios. 
For example, a person in the height of her carrier might be most interested in keeping the managerial position.
In contrast, during a recession, the focus might be in finding a new job regardless of the hierarchical position, hence the validity of both projections.

\subsection{Comparison of ranked lists}
\begin{figure}[h]
\centering
\includegraphics{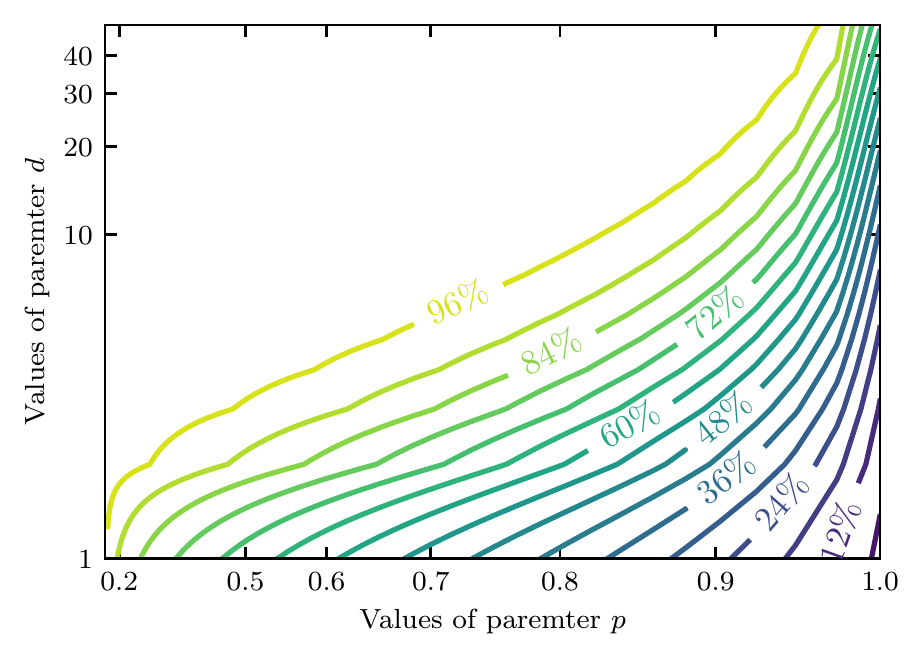}
\caption{Percentage of contribution of the top $d$ elements in the lists to the value of $RBO$~\eqref{eq:rbo} for different values of parameter $p$. 
The axis scale for $p$ is distorted in order to emphasise the the influence of $p$ values for longer lists. }
\label{fig:weight_rbo}
\end{figure}

The next step in comparing the proposed similarity measures is analysing the resulting ranked lists.
This can be done using measures like Kendall's tau or a more complete measure called ranked based overlap~\cite{10.1145/1852102.1852106}.
For two (possibly infinite long) ranked lists $L_1$ and $L_2$, RBO is defined as:
\begin{equation}
RBO(L_1,L_2,p) = (1-p) \sum_{d=1}^{\infty}p^{d-1} A_d\text{ with } 0<p<1,
\label{eq:rbo}
\end{equation}
where $d$ is the depth to which the lists are analysed, $X_d = | L_1 \cap L_2 |$ is the cardinality of the intersection of both lists at depth $d$ and $A_d = \frac{X_d}{d}$ is called \emph{agreement}, i.e.\ proportion of the lists that have the same elements.
The value $p$ determines the relative weight of the first $d$ ranks to the overall value of~\eqref{eq:rbo}.
For limit case, $p\to1$~\eqref{eq:rbo} returns unbounded overlap intersection~\cite{10.5555/644108.644113}.
The contribution of the first $d$ elements for a particular $p$ can be determined as~\cite[eq (21)]{10.1145/1852102.1852106}
\begin{equation}
W(p,d) = 1-p^{d-1}+\frac{d(1-p)}{p}\left( \ln \frac{1}{1-p} - \sum_{d=1}^{\infty}\frac{p^i}{i}\right).
\label{eq:weight_rbo}
\end{equation}
\figurename~\ref{fig:weight_rbo} provides a visualisation of the relation~\eqref{eq:weight_rbo}.
It can be seen that as the length $d$ of the lists increases, the parameter $p$ should be fine tuned in order to obtain the desired contribution of the top $d$ components to the overall RBO value.

\begin{figure}[h]
\centering
\includegraphics{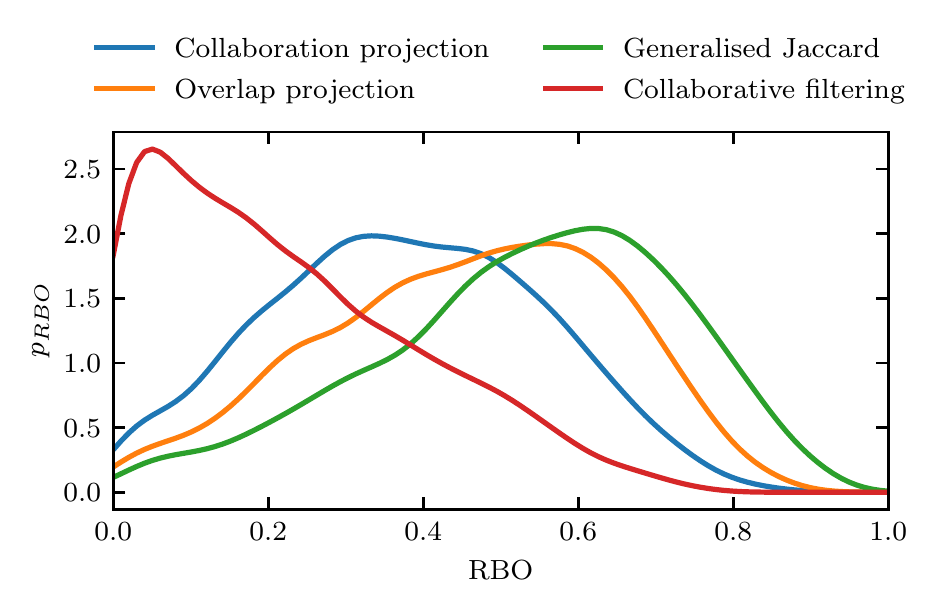}
\caption{Estimated probability density of RBO values~\eqref{eq:rbo} over all 2941 \gls{esco} occupations calculated with $p=0.9$.
The RBO values were caluclated between Nesta occupation similarity and various bipartite graph projections.}
\label{fig:rbo_hists}
\end{figure}

When comparing occupation similarity ranks, setting $p=0.9$ puts 96\% of the RBO value contributed by the top 20 most similar occupations in the corresponding lists.
In the case of \gls{esco}, there are 2941 occupations and the same number of ranked lists for each similarity measure. 
We calculate the RBO values for each occupation between the lists ranked with Nesta weights and weights obtained by various bipartite projections.
The result are 2941 RBO values for each projection, whose histograms are shown in \figurename~\ref{fig:rbo_hists}. 

The distribution of the RBO values show that the Nesta occupation similarity provides rankings that are quite consistent with the ones provided with the generalised Jaccard projection~\eqref{eq:w_jacc}.
This is expected since, among other steps, Nesta uses a cosine distance on an occupation partition that is very similar to the one used here.

\subsection{Validation on actual job transfers in Slovenia}  
When addressing the issues of job similarity, it is not possible to perform validation in its basic sense, since there is no unified true value for job similarity. 
To some extent, this can be resolved by observing how certain similarity measures correspond to the actual observed job transfers.
In this case, this is done using micro-data records from the Slovenian Statistical office, which include every job transfer occurring in the economy.
For the period from 2012 onwards, over 450,000 job transitions are recorded.
This gives a unique insight into the actual occupation transfers.

A way of validating the proposed similarity measure is through observation of job transfers~\cite{10.1145/963770.963772}.
One expects the transfers to occur more frequently among occupations that are ``closer'' (more similar).

The \gls{sors} data are represented using a set of 4-level codes of \gls{isco} $\mathcal{I}$, comprising 430 different occupations.
The similarity measures between \gls{esco} occupations thus have to be mapped onto \gls{isco} codes.
The mapping from the set of \gls{esco} occupations $\mathcal{O}$ to 4-level \gls{isco} codes $\mathcal{I}$
\begin{equation}
f: \mathcal{O}^{2941} \to \mathcal{I}^{430}
\label{eq:esco_isco_map}
\end{equation}
is a non-injective surjective function\footnote{Multiple occupations $o\in\mathcal{O}$ map to a same occupation $o^{\mathcal{I}}\in\mathcal{I}$.} that is provided partially within the \gls{esco} structure and partially done by the public employment service of Slovenia\footnote{This mapping is available in the supplementary materials.}.
As a result, the distance matrix $D^{\mathcal{O}}$ calculated for \gls{esco} codes is mapped to the $D^{\mathcal{I}}$ of the \gls{isco} codes as
\begin{equation}
d^{\mathcal{I}}_{mn} = \max\{d^{\mathcal{O}}_{ij}|f(i)=m, f(j)=n\}.
\label{eq:isco_d}
\end{equation}

Having the similarity measure matrix~\eqref{eq:isco_d}, the number of transfers between pairs of occupations $o_m$ and $o_n$ are binned in the job similarity value being tested $d^{\mathcal{I}}_{mn}$ and plotted as a histogram.

\begin{figure}[h]
\centering
\includegraphics{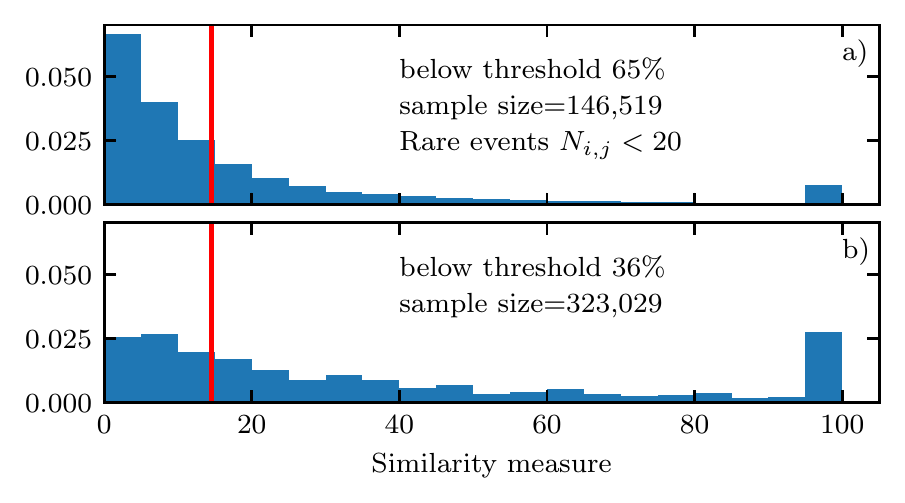}
\caption{Distribution of the collaborative projection weights~\eqref{eq:recom}.}
\label{fig:collab}
\end{figure}

\begin{figure}[h]
\centering
\includegraphics{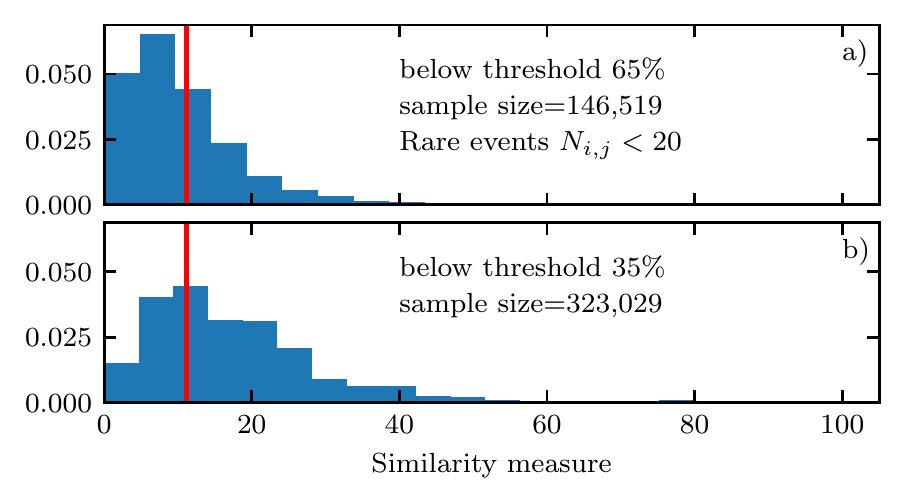}
\caption{Distribution of the generalised Jaccard projection weights~\eqref{eq:w_jacc}.}
\label{fig:weight}
\end{figure}

\begin{figure}[h]
\centering
\includegraphics{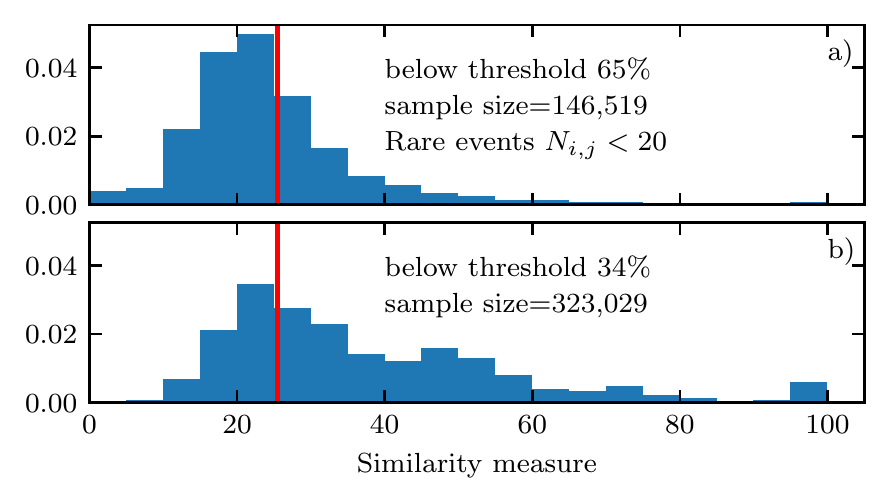}
\caption{Distribution of the Nesta calculated weights.}
\label{fig:nesta}
\end{figure}

\figurename s~\ref{fig:collab}~to~\ref{fig:nesta} show the histogram distribution of $d^{\mathcal{I}}_{mn}$ similarities calculated for generalised Jaccard distance~\eqref{eq:w_jacc}, collaborative filtering projection~\eqref{eq:recom} and weights from Nesta's calculation respectively.
Since each of these projections provides similarity measure values on a different scale, the subsequent comparison is done using the following transformations:
\begin{enumerate}
\item If the number of observed transitions between occupations $o_m$ and $o_n$, $N_{m,n}$, is below 20, the transition is regarded as rare;
\item The $d^{\mathcal{I}}_{mn}$  values are normalised, so that all values of $d^{\mathcal{I}}_{mn}$ up to the \engordnumber{98} percentile are mapped in the interval $[0,1]$.
The values above the \engordnumber{98} percentile are fixed at 1; and
\item A threshold value for each projection is selected so that 65\% of the rare job transitions have a similarity measure below this value.
\end{enumerate}
In such a way, the resulting histogram plots can be compared in a more direct manner.

The three similarity measures are treated as classification models for classifying occupation transfers into rare with $N_{m,n}<20$ and common with $N_{i,j}\ge 20$.
The latter class is chosen to be positive.
By the choice of the threshold, the true negative rate of the models is 65\%.
All three similarity measures show a similar performance, i.e. for common occupation transfers, an almost identical majority of the similarity weights is above the selected threshold.
RBO between the collaboration filtering projection~\eqref{eq:recom} and the Nesta weights show quite low correspondence (see \figurename~\ref{fig:rbo_hists}).
However the calculated similarity measure values seem to be inline with the observed transfers in the micro-data.
This is additional evidence that different occupation similarity measures capture different dynamics in the society.

\subsection{Discussion on the results}
\label{sec:discussion}
\subsubsection*{Pairs of occupations vs. observed number of transfers between them}
The assessment of occupation similarity measures can be done using two different scenarios: through the number of observed occupation transfers or through occupation pairs.
There is a subtle but important distinction between these two approaches. 

The first approach takes into account the observed number of transitions between occupations, as in the above analysis.
Consequently, the results do not take into account information about the similarity values for occupation pairs for which there were no observed transitions. 
As a result, in \figurename s~\ref{fig:collab}--\ref{fig:nesta}, histograms a) are dominated by occupations that have almost 20 transitions.

In the second scenario, the occupations with a large number of transfers are treated as equally as important as those without or with a very small number of observed transfers.
The number of transfers is used only to determine whether a particular pair of occupations should be treated as a rare transfer or not.
As a result, the analysis of rare occupations becomes more dominant.
The trade-off is seen on the side of the occupation pairs with a large number of transfers, whose importance is relatively diminished.

For completeness of the analysis, the results of the second scenario are presented in Appendix~\ref{app:pairs}.
Those results also confirm the viability of the various occupation similarity measures.

\subsubsection*{Accuracy of the results}
At first glance, one might see the results as being contradictory.
On one side, \figurename~\ref{fig:rbo_hists} shows that projections from either binary or weighted bipartite graphs result in quite different ordering.
This does not mean that some of the projections are underperforming compared to Nesta, but are simply showing that the provided ranks are not that similar to the ones provided by Nesta.
On the other side, the frequentist analysis shown in \figurename s~~\ref{fig:collab}--\ref{fig:nesta} state that all proposed similarity measures are capable of distinguishing ``rare'' job transfers from the ``more'' expected ones.

\begin{figure}[h]
\centering
\includegraphics{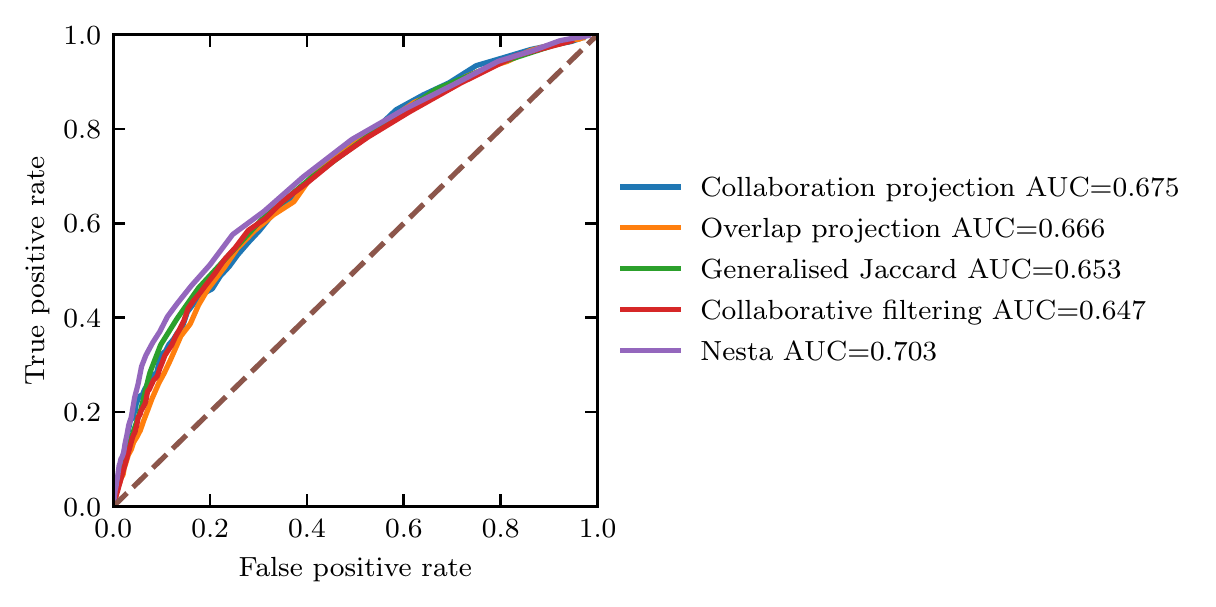}
\caption{ROC curves using different occupation similarity measures.}
\label{fig:roc}
\end{figure}

Using these results, it is possible to draw ROC curves~\cite{10.1177/2192568218778294} for each projection, as shown in \figurename~\ref{fig:roc}.
ROC curves are calculated from the empirical distributions derived from the histograms shown in \figurename s~~\ref{fig:collab}--\ref{fig:nesta} by varying the threshold.
The histograms a) in \figurename s~~\ref{fig:collab}--\ref{fig:nesta} represent negative cases and the positive cases are those with $N_{m,n} \ge 20$, i.e. histograms b) in \figurename s~\ref{fig:collab}--\ref{fig:nesta}.
True positive rate is the proportion of cases above the current threshold in histograms b); and, false positive rate is the same proportion for the same threshold in histograms a).
All of the above similarity measures have a similar performance with the area under the curve of between 65--70\%.
One may argue that this might be considered as an indication of underperforming.
There are three main reasons for this.
First, when doing the frequentist analysis, only the last position was taken into account. 
This means that the complete job history of each person was not taken into account.
Second, the the mapping from \gls{esco} to \gls{isco} codes~\eqref{eq:isco_d} in some cases introduces additional errors.
Finally, there are the actual misclassifications at the registry level.
Despite the highly granular occupation structures (or maybe because of it), there are cases when persons' occupation registrations are not accurate, mainly due to human decision making in the HR departments.

\section{Conclusion}
The performed analysis conveys three key results.
First, it is possible to use various bipartite projections in order to calculate logical occupation similarity values using only an abstract occupation-skills mapping like ESCO.
Second, different projections offer different insights into possible career evolution, hence different similarity values.
Therefore, it is possible to define scenarios specifying when to use a particular projection based on the person's career wishes or other factors.
Finally, the applicability of the calculated similarity measures are validated on a micro-level data set of more than 450,000 transitions, showing that the results are consistent with the observed real-world job transfers.

Apart from the applicability of the bipartite projections for the occupation similarity problem, all of the listed approaches have efficient computational implementation.
Therefore, it is possible to introduce a whole palette of measures that provide different rankings thus tailoring the behaviour of the output.
As shown in this analysis, even for projections that provide quite different rank overlap results, real-world data show that a substantial number of such transitions are observed, hence their viability. 
So, using a single general occupation similarity measure might be a limiting factor compared to the flexibility of having explainable measures offering different career evolution pathways.

\subsection*{Code availability}
The numerical implementation of the analysed projections and corresponding data sets are available at~\url{https://repo.ijs.si/pboskoski/bipartite_job_similarity_code}.

\section*{Acknowledgements}
The authors acknowledge the research core funding No.\ P2-0001 financially supported by the Slovenian Research Agency.
The authors also acknowledge the funding received from the European Union's Horizon 2020 research and innovation programme project HECAT under grant agreement No.\ 870702.
The micro-data employer-employee datasets were provided by the \glsdesc{sors}.

 \appendix
\section{Analysis of occupation pairs}
\label{app:pairs}

As stated in Section~\ref{sec:discussion}, the second scenario of evaluating the occupation similarity ranking is based only on the occupation pairs, i.e. it does not take into account the number of observed transfers.
The only point where the number of transfers is used is for dividing the occupation pairs in two groups: rare and common.
As in the analysis from Section~\ref{sec:discussion}, the threshold is set at 20 transfers.

The analysis in Section~\ref{sec:discussion} contains almost 44,000 distinct occupation pairs for which a job transition was observed.
On the other hand, in Slovenia there are 430 distinct \gls{isco} codes and taking into account only pairs with different occupations results in 184,470 occupation pairs for which a similarity can be calculated\footnote{Similarities were calculated for all \gls{esco} occupation pairs, i.e. $2941^2$ pairs.}.

\begin{figure}[h]
\centering
\includegraphics{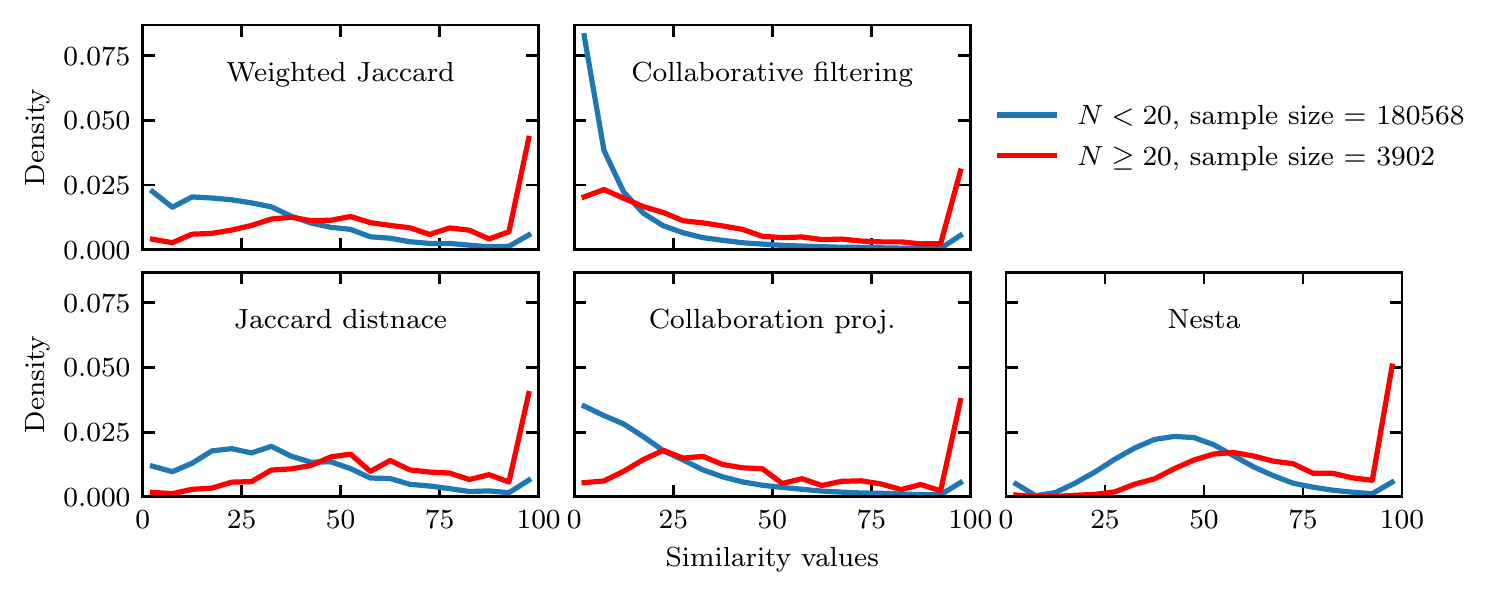}
\caption{Density histograms of occupation similarity measures taking into account only occupation pairs.
A threshold of 20 transfers is used only to divide the data set into rare and common pairs, as in Section~\ref{sec:ranking}.}
\label{fig:pairs}
\end{figure}

The most visible change is the shape of the rare occupation pairs histogram, drawn with blue lines in \figurename~\ref{fig:pairs}.
These histograms now contain information from occupation pairs that did not have any observed job transfers in the last 10 years, which amounts to roughly $\frac{3}{4}$ of all possible occupation pairs.
Despite this, the results show that the calculated similarity values are low, i.e. all blue-line histograms are left skewed.

\begin{figure}[h]
\centering
\includegraphics{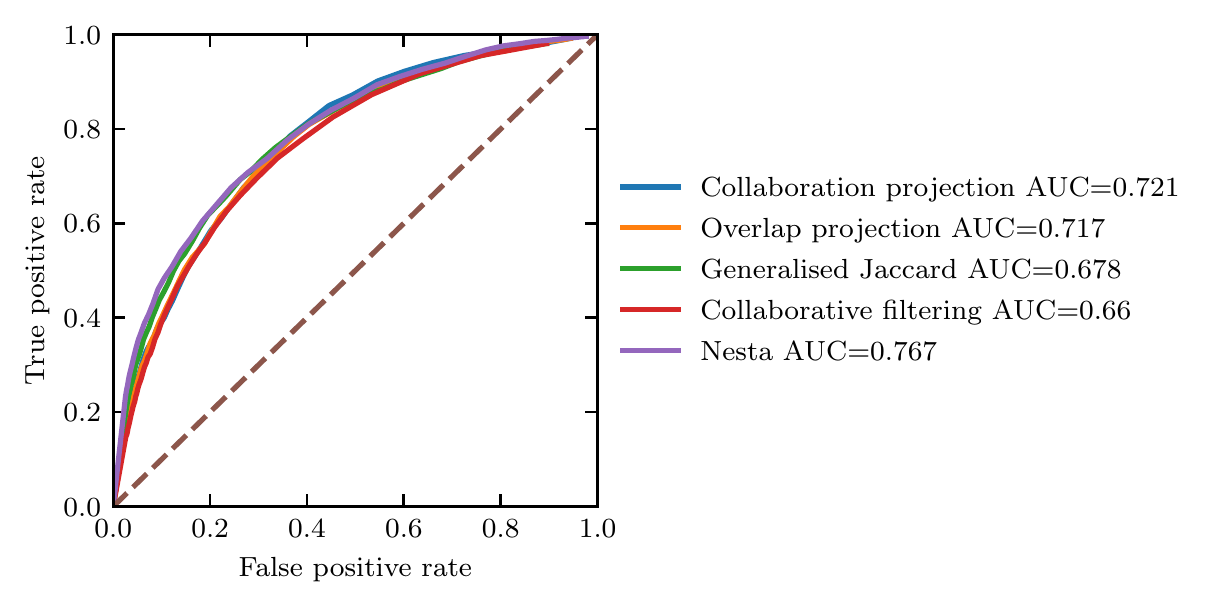}
\caption{ROC curves calculated from distributions in \figurename~\ref{fig:pairs}.}
\label{fig:roc_pairs}
\end{figure}

Furthermore, the ROC curves calculated using these histograms are shown in \figurename~\ref{fig:roc_pairs}.
The results are comparable to those shown in \figurename~\ref{fig:roc}.
This confirms that despite providing a different ranking, all of the analysed similarity values have a comparable classification performance. 

\printbibliography

\end{document}